\documentstyle[aps,psfig]{revtex}
\begin{document}
\newcommand{\Prd}{Phys. Rev D}
\newcommand{\Prl}{Phys. Rev. Lett.}
\newcommand{\Pl}{Phys. Lett.}
\newcommand{\Cqg}{Class. Quantum Grav.}
\newcommand{\Fstar}{\!\stackrel{*}{F}{\!\!}}
\newtheorem{lemma}{Lemma}
\renewcommand{\thelemma}{\Alph{lemma}}
\newcommand{\beq}{\begin{equation}}
\newcommand{\eeq}{\end{equation}}
%\preprint{ICI-EFEI-002/99}
\draft
\title{A nongravitational wormhole}
\author{F. Baldovin, M. Novello%
\thanks{Electronic mail: \tt novello@lafex.cbpf.br},  
S. E. Perez Bergliaffa, and 
J. M. Salim}
\address{Centro Brasileiro de Pesquisas F\'{\i}sicas \\
Rua Dr.\ Xavier Sigaud 150, Urca 22290-180 Rio de Janeiro, RJ -- Brazil} 
\date{\today}
\renewcommand{\thefootnote}{\fnsymbol{footnote}}
%\twocolumn[
%\hsize\textwidth\columnwidth\hsize\csname@twocolumnfalse\endcsname 
\maketitle

\begin{abstract}
\hfill{\small\bf Abstract}\hfill\smallskip
\par
Using the effective metric formalism for photons in a nonlinear electromagnetic theory, we show that a certain field configuration in Born-Infeld electromagnetism in flat spacetime can be interpreted as an ultrastatic spherically symmetric 
wormhole. We also discuss some properties of the effective metric that are valid for any field configuration.
\end{abstract}

%\pacs{PACS numbers:  }
%\smallskip\mbox{}]
%\renewcommand{\footnotemark}{\fnsymbol{\thefootnote}}
%\renewcommand{\thefootnote}{\fnsymbol{footnote}}
%\footnotetext[1]{Electronic mail: \tt novello@lafex.cbpf.br}
%\addtocounter{footnote}{1}
\renewcommand{\thefootnote}{\arabic{footnote}}

\section{Introduction}

There has been an increasing interest in recent years in earth-bound systems 
that can mimic gravitational and cosmological phenomena. These systems are of a 
very different nature: superfluid $^3$He-A and high-energy superconductors 
\cite{vol}, supersonic acoustic flows \cite{u1,u2,visser}, flowing dielectric 
fluids \cite{leon}, and Bose-Einstein condensates \cite{garay}. A common feature 
of some of these systems is that 
the propagation of disturbances in a fixed flat background is described by the 
equation of motion of a massless scalar field in a curved ``effective'' spacetime. The metric 
of this curved spacetime depends on the physical system under consideration. In this 
reformulation in terms of Lorentzian geometry it can be shown that, under 
certain hypothesis, 
black-hole analogues are possible in some of the abovementioned systems. There 
are even some chances to probe semiclassical quantum gravity in the laboratory, 
through the Hawking effect \cite{u1,u2,visser}. We would like to study here yet 
another system in which an effective metric can be introduced to describe the 
propagation of waves. It is a well-known fact that Maxwell's Lagrangian acquires 
nonlinear correction terms due to vacuum polarization. 
Novello {\em et al} 
\cite {nov1} showed recently that
in {\em any} nonlinear electromagnetic (EM) theory  
photons, in the geometrical optics approximation, {\em do not} propagate along null geodesics of the background geometry. 
They propagate instead along null geodesics of an effective geometry, which 
depends on the nonlinearities of the dynamics of the theory. 
This result was derived by Pleba\'nsky in the case of Born-Infeld 
electrodynamics in 
1968 \cite{pleb}, and was in \cite{nov1} extended to encompass any nonlinear 
electromagnetic theory. The propagation of photons in nontrivial QED vacua
has also been studied 
in \cite{Latorre,Dittrich,Shore}. 
Very often these analysis have led to unexpected results.  
As examples, let us mention the possibility 
of {\em faster-than-light} photons \cite{ftlp}, and the existence of closed 
timelike curves due to nonlinear EM effects \cite{nov2}.
Morover, it was shown in \cite{nov1} that the the motion of light through a 
nontrivial vacuum  
is equivalent to 
that of photons propagating according to Maxwell's theory in a nonlinear medium. The medium induces 
modifications on the equations of motion 
which are identical to those produced by the nonlinearities of the field.
We would like to address here some general features of this effective geometry, 
and also to show an example of a specific effective spacetime for photons: an 
electromagnetic wormhole. 

The structure of the paper is the following. In Section \ref{geometry} we review 
the derivation of the effective geometry for photons in nonlinear 
electrodynamics, and we point out some of its general features. In 
Section \ref{wormhole} we show how a wormhole can be constructed out of the 
effective geometry in the case of Born-Infeld electrodynamics, and we analize 
its properties. We close with a discussion of the results.

\section{Effective geometry for photons}
\label{geometry}

In this section we present the method of the effective geometry \cite{nov1}. 
Let us start with a Lagrangian density that is a function of $F=F_{\mu\nu}F^{\mu\nu}$ 
only
\footnote{ The more general case in which ${\cal L} = {\cal L}(F,F^*)$, 
where $F^*\equiv F_{\mu\nu}F^{*\mu\nu}$ and $F^{*\mu\nu}$ is the dual of 
$F^{\mu\nu}$, 
can be found in \cite{nov1}.}. We will follow Hadamard's technique \cite{had} to study the 
propagation of the discontinuities of the EM field. Let $\Sigma$ a surface of 
discontinuity
of $F_{\mu\nu}$, with equation $\Sigma (x^\mu ) = {\rm const}$. We assume that 
the EM tensor is continuous when crossing $\Sigma$, but its derivative has a 
finite discontinuity. These assumptions can be stated as

\beq
[F_{\mu\nu}]_\Sigma = 0, ~~~~~~~~~~~~~~~[\partial_\lambda F_{\mu\nu}]_\Sigma = 
f_{\mu\nu}k_\lambda .
\label{disc}
\eeq
The operation $[~~]_\Sigma$ on a tensor $J$ is defined by
\beq
[J]_\Sigma \equiv {\rm lim}_{\;\delta\rightarrow 0^+} \left( J|_{\Sigma + \delta} 
- J|_{\Sigma - \delta}\right),
\eeq
and represents the discontinuity of the tensor $J$ through $\Sigma$. The tensor 
$f_{\mu\nu}$ is the discontinuity of the field, and $k_\lambda = 
\partial_\lambda \Sigma$ is the propagation vector.

If we apply $[~~]_\Sigma$ to the nonlinear equations of motion
\beq
(\sqrt{-\gamma}\;{\cal L}_F F^{\mu\nu})_{,\nu}= 0,
\label{nlmax}
\eeq
(${\cal L}_F$ is the derivative of the Lagrangian density w.r.t. $F$,
$\gamma_{\mu\nu}$ is the metric of Minkowski's spacetime 
in an arbitrary coordinate system, and $\gamma$ is the corresponding determinant) we obtain
\beq
{\cal L}_F f^{\mu\nu} k_\nu + 2 {\cal L}_{FF} \xi F^{\mu\nu}k_\nu = 0,
\label{disc1}
\eeq
where $\xi \equiv F^{\alpha\beta}f_{\alpha\beta}$. 
The same operation on the 
cyclic identity yields

\beq
f_{\mu\nu}k_\lambda + f_{\nu\lambda}k_\mu + f_{\lambda\mu}k_\nu = 0.
\nonumber
\eeq
This equation can be rewritten after contraction with 
$k_\alpha\gamma^{\alpha\lambda}F^{\mu\nu}$. The result is 

\beq
\xi k_\mu k_\nu \gamma^{\mu\nu} + 2 F^{\mu\nu} f _\nu ^\lambda k_\lambda k_\mu = 
0.
\label{disc2}
\eeq
We can eliminate the tensor 
$f_{\mu\nu}$ from Eqs.(\ref{disc1}) and (\ref{disc2}). If we assume that $\xi$ and ${\cal L}_F$ are nonzero, we get

\begin{equation}
\left({\cal L}_F\gamma^{\mu\nu} - 4{\cal 
L}_{FF}F^{\mu\alpha}F_\alpha{}^\nu\right) 
k_\mu k_\nu = 0.
\label{gww4}
\end{equation}
This expression 
suggests 
that the 
self-interaction of the field $F^{\mu\nu}$ can be interpreted
as a modification on the spacetime metric 
$\gamma_{\mu\nu}$ which is described by the effective geometry  
\begin{equation}
g^{\mu\nu}_{\rm eff} = {\cal L}_{F}\,\gamma^{\mu\nu}  - 
4\, {\cal L}_{FF} \,{F^{\mu}}_{\alpha} \,F^{\alpha\nu}.
\label{geffec}
\end{equation}
Note that only in 
the particular case of linear electrodynamics 
the discontinuities of the  electromagnetic field 
propagate along null paths in the Minkowski background.

In the derivation of Eq.(\ref{geffec}) we assumed that $\xi$ and ${\cal L}_F$ do 
not have zeros. It can be proved (see \cite{nov1}) that photons such that $\xi 
=0$ propagate along geodesics of the Minkowskian geometry. To analyse the other 
restriction, note that ${\cal L}_F$ is a function of the coordinates $x^\mu$ 
through the EM tensor $F_{\mu\nu}(x^\mu)$. If the theory given by  
${\cal L}$ and the field configuration under study are such that ${\cal L}_F$ 
has one or more zeros for certain values $x^\mu_0$ of the coordinates, 
Eq.(\ref{disc1}) will in general not be satisfied in those $x^\mu_0$. However, the 
equation is valid for points that are arbitrarily close to $x^\mu_0$, and hence 
the effective metric tensor is everywhere well-defined except for these  zeros.

The general expression of the effective geometry can be
equivalently written in terms of the vacuum expectation value  
of the energy-momentum tensor, given by 
\begin{equation}
T_{\mu\nu} \equiv \frac{2}{\sqrt{-\gamma}} 
\,\frac{\delta\,\Gamma}{\delta\,\gamma^{\mu\nu}},
\nonumber
\end{equation}
where $\Gamma$ is the effective action
\begin{equation}
\Gamma \doteq \int \,d^{4}x \sqrt{-\gamma}\,{\cal L}.
\nonumber
\end{equation}
In the case of one-parameter Lagrangians, ${\cal L}={\cal L}(F),$ we obtain
\begin{equation}
T_{\mu\nu} = - 4 {\cal L}_{F}\, {F_{\mu}}^{\alpha} \, F_{\alpha\nu} - 
{\cal L}\,\eta_{\mu\nu},  
\nonumber
\end{equation}
where we have chosen a Cartesian coordinate system in 
which $\gamma_{\mu\nu}$ reduces to $\eta_{\mu\nu}.$ 
In terms of this tensor the effective geometry Eq.(\ref{geffec}) 
can be re-written as%
\footnote{ For simplicity, we will denote the effective metric 
as $g^{\mu\nu}$ instead of $g^{\mu\nu}_{\rm eff}$ from now on.} 
\begin{equation}
\label{gT}
g^{\mu\nu}=\left({\cal L}_F+\frac{{\cal L}\,{\cal L}_{FF}}{{\cal 
L}_F}\right)\eta^{\mu\nu} 
+\frac{{\cal L}_{FF}}{{\cal L}_F}T^{\mu\nu}.
\label{emmet}
\end{equation}
This equation shows that
the stress-energy distribution of the field is the true responsible for 
the departure of the effective geometry
from its Minkowskian form. It is also seen that 
for $T_{\mu\nu} = 0$, the conformal 
modification in Eq.(\ref{gT}) clearly leaves the photon paths unchanged.

At this point several remarks are in order:

\begin{itemize}

\item Note that there are two metric tensors present in the problem. One is the 
Minkowskian metric $\gamma_{\mu\nu}$. In the absence of forces, ordinary matter moves on geodesics of this 
metric. The second is the effective metric $g_{\mu\nu}$, and only influences the motion of the photons.

\item The properties of the effective geometry depend on the specific theory 
under consideration (through ${\cal L}$ and its derivatives), and also on the 
specific field configuration. The latter is determined by solving the nonlinear 
Eqns.(\ref{nlmax}) together with $\partial_{[\mu}F_{\nu\rho ]}= 0$.

\item The effective geometry defined by Eq.(8) admits an 
everywhere nonzero conformal factor. 
This freedom is due to the fact that the 
effective geometric structure is valid only for photons. It follows that only 
concepts that do not depend on the choice of the conformal factor 
have a definite meaning in this framework.

\item The use of the effective geometry is not mandatory. 
Some results can be also obtained by other methods, although these require an average over the polarization states \cite{Dittrich}.

\item Although we shall restrict here to flat backgrounds, the effective 
geometry can and must be used in the presence of curved backgrounds. In this case, the flat spacetime metric $\eta_{\mu\nu}$ is replaced by the curved spacetime metric $g_{\mu\nu}$. The latter is consistently determined by Einstein's equations.
We must point out that if the change in the path of the photons described by the effective metric is not taken into account, the resultant description of the spacetime under consideration is incomplete. For instance, it was shown in 
\cite{nov3} that an apparently ``regular'' black hole generated by a certain nonlinear 
electrodynamics 
is actually singular, the singularities in the effective geometry being associated to 
the nonlinearities of the theory.

\end{itemize}

Let us now turn to a far-reaching analogy between nonlinear EM on one side and 
linear EM in the presence of dielectrics on the other.
In \cite{nov1} was shown that it is possible to describe the propagation of waves governed 
by Maxwell 
electrodynamics inside a medium in terms of a modification of the 
underlying spacetime geometry using the framework developed above.
The electromagnetic field inside a material medium can be represented by two 
antisymmetric
tensors, 
the EM field $F_{\mu\nu}$ and the polarization $P_{\mu\nu}$, 
which in the absence of sources obey 
the equations
\beq
\partial^\nu P_{\mu\nu} = 0, ~~~~~~~~~~\partial^\nu F^{*}_{\mu\nu} = 0.
\eeq
From these equations and Eq.(\ref{nlmax}) we see that the nonlinear theory in 
vaccum is equivalent to Maxwell's theory in an isotropic medium characterized by
an electric susceptibility $\epsilon$ and a magnetic permeability $\mu$ given by
\beq
\epsilon = {\cal L}_F ~~~~~~~~~ \mu = \frac{1}{{\cal L}_F}
\eeq
(see \cite{nov1} for details).
Therefore, the simple class of effective Lagrangians ${\cal L} = {\cal L}(F)$ 
may be used as a convenient description of Maxwell theory 
inside isotropic media; 
conversely, results obtained in the latter context 
can as well be similarly restated in the former one
\footnote{ Let us remind the reader that $\epsilon$ and $\mu$ are functions of the 
EM field, and so their derivatives are discontinuous on the wavefront.}.
%In particular, 
%for electrostatic fields inside 
%an isotropic dielectric medium the effective geometry can be rewritten as
%\begin{equation}
%\label{gef22}
%g^{\mu\nu} = \epsilon\, \eta^{\mu\nu} - \frac{\epsilon\rq}{E} \left( 
%E^{\mu}\,E^{\nu} - E^2\, \delta^{\mu}_t \,\delta^{\nu}_t \right),
%\end{equation}
%where $E^2 \equiv -\,E_{\alpha}\,E^{\alpha} > \,0.$
%In other words,
%\begin{eqnarray}
%\label{gef223}
%g^{tt} &=& \epsilon  + \epsilon' E \\
%\label{gef224}
%g^{ij} &=& -\,\epsilon\,\delta^{ij} - \frac{\epsilon'}{E}\, E^i\,E^j .
%\end{eqnarray}
%This shows that the discontinuities of the electromagnetic field 
%inside a nonlinear dielectric medium propagate along null cones 
%of an effective geometry which depends on the characteristics 
%of the medium.  

\section{A wormhole for photons}
\label{wormhole}

In this section we show that there exists a field configuration in Born-Infeld 
nonlinear EM theory that generates an effective geometry which is a wormhole for photons. 
We shall restrict here to a spherically symmetric and static effective 
metric. It can be proved (see Appendix) that the effective metric 
tensor expressed in spherical coordinates of the Minkowskian background takes the form

\beq
ds^2_{\rm eff} = \Phi^{-1} (dt^2 - dr^2) - \Psi ^{-1} r^2 d\Omega^2,
\label{effmet}
\eeq
where 
\beq
\Phi = {\cal L}_F - 4 {\cal L}_{FF} A^2,
\nonumber
\eeq
\beq
\Psi = {\cal L}_F + 4 {\cal L}_{FF} B^2 r^4 \sin^2\theta ,
\eeq
and the only nonzero components of the EM tensor are $A\equiv F^{tr}$ and $B\equiv F^{\theta\varphi}$. From the nonlinear Eqs.(\ref{nlmax}) 
and the cyclic identity, we get

\beq
A = \frac{\alpha}{{\cal L}_F r^2}, ~~~~~~~~~~~ B = \frac{\beta}{r^4 sin\theta},
\eeq
with $\alpha$ and $\beta$ arbitrary constants.
From these equations, it is seen that the electromagnetic tensor $F_{\mu\nu}$ 
must satisfy 
\beq
F = \frac{2}{r^4} \left(\beta^2-\frac{\alpha^2}{{\cal L}_F^2}\right).
\label{const}
\eeq
We now adopt as an example Born-Infeld electrodynamics \cite{bi}. The Lagrangian density 
is given by

\beq
{\cal L} = b^2\left( 1-\sqrt{1 + \frac{F}{2b^2}}\;\right).
\nonumber
\eeq
Note that ${\cal L}_F$ and ${\cal L}_{FF}$ are nonsingular functions of $r$ for 
this Lagrangian.
The constraint Eq.(\ref{const}) gives in this case 

\beq
F = 2b^2\frac{\beta ^2 - 16\alpha^2}{b^2r^4 + 16 \alpha ^2}
\eeq
and the metric functions are
\beq
\Phi = -\frac{1}{4b^2 r^2}\frac{(r^4 
b^2+16\alpha^2)^{3/2}}{\sqrt{b^2r^4+\beta^2}},
\eeq
\beq
\Psi = 
-\frac{1}{4b^2r^4}\frac{\sqrt{b^2r^4+16\alpha^2}(b^4r^8-16\alpha^2\beta^2)}{(b^2 
r^4+\beta^2)^{3/2}}.
\eeq
We see that the function $\Psi$ has a zero, and as a result, the metric coefficient 
$\Psi^{-1}r^2$ will be singular.
From now on we restrict to the case in which only the electric field is nonzero, 
{\em i.e.} $\beta = 0$. In this case, the metric functions are

\beq
\Phi = -\frac{(b^2r^4+16\alpha^2)^{3/2}}{4b^3r^6} , 
\eeq
\beq
\Psi = -\frac{\sqrt{b^2 r^4 + 16 \alpha ^2}}{4br^2} .
\eeq

Due to the freedom 
in the conformal factor of the definition of the effective metric (see remarks above) we shall 
analyze the features of the metric

\beq
ds^2 = dt^2 - dr^2 - \Delta (r) d\Omega ^2,
\label{metric1}
\eeq
where $\Delta (r) \equiv r^2\Phi /\Psi$ is given by

\beq
\Delta = \frac{b^2r^4+16\alpha^2}{b^2r^2}=r^2+\frac{r_{\rm th}^4}{r^2},
\eeq
with $r_{\rm th}\equiv 2\sqrt{|\alpha|/b}$.
Note that the metric is singular in $r=0$. However, this divergence is an artifact of the coordinate system that comes from the behaviour of the spherical coordinates at $r=0$.

From the function $\Delta (r)$ we can calculate the area of the 
2-surface $t={\rm const}$, $r={\rm const}$, which is given by $A^{(2)} = 
4\pi\Delta (r)$.
The function $A^{(2)}$ 
has a minimum at $r=r_{\rm th}$, and does not have zeros. For 
large values of $r$, $A^{(2)}\rightarrow 4\pi r^2$ and the metric goes into the flat 
spacetime metric. At $r=0$,  $A^{(2)}$ diverges due to the singularity of the metric. This analysis shows that the effective geometry for 
the photons is in fact a wormhole \cite{vishoch}, with a throat located at $r_{\rm 
th}$, and two different asymptotic regions.

The wormhole structure may also be displayed by making a coordinate change 
to write the metric in the standard form. By setting

\beq
\rho^2\equiv r^2 + \frac{r_{\rm th}^4}{r^2};
\nonumber
\eeq
and inverting this relation, we can write the metric Eq.(\ref{metric1}) in the form

\beq
ds^2 = dt^2 - \frac{1}{1-\frac{b_\pm (\rho_\pm)}{\rho_\pm}} d\rho_\pm ^2 - \rho_\pm^2 d\Omega ^2,
\label{awhmet}
\eeq
with

\beq
b_\pm(\rho_\pm) = \frac{2\rho_{\rm th}^4-\rho_\pm ^4\pm \rho_\pm 
^2\sqrt{\rho_\pm ^4-\rho_{\rm th} ^4}}{\rho_\pm ^3\pm\rho_\pm \sqrt{\rho_\pm 
^4-\rho_{\rm th} ^4}},
\eeq
and $\rho_{\rm th}\leq\rho_\pm<\infty$. The ``$+$'' patch covers the region 
$r_{\rm th}\leq r<\infty$, and the ``$-$'' patch, the region $0<r\leq r_{\rm 
th}$. 
The functions $b_\pm (\rho_\pm)$ satisfy the requirements needed in order to have a wormhole geometry \cite{vishoch}, with a throat located at $\rho_{\rm th}\equiv \sqrt{2}\,r_{\rm th}$. 
Because two different shape functions $b(\rho)$ are needed, the wormhole is asymmetric under the interchange $\rho_+\leftrightarrow\rho_-\;$. Note also that only the region covered by the ``+'' patch is asymptotically flat.

The same analysis can be carried out for the magnetic case, {\em i.e.} $\alpha = 
0$, $\beta \neq 0$. It follows that the effective metric also describes a 
wormhole, with a throat located at $r_{\rm th} =\sqrt{|\beta |/b}$, and 
asymptotic regions as in the electric case.

To analyze the motion of the photons in the effective geometry, we can use the effective potential. In the case of a static and spherically symmetric spacetime, $L$ and $E$ represent constants of motion along a geodesic path:

\beq
g_{\varphi\varphi} \dot{\varphi}=L~~~~~~~~~~g_{tt}\dot{t}=E,
\eeq
where the dot indicates derivation w.r.t. an affine parameter, and $\theta=\pi /2$. If we consider null curves we obtain

\beq
g_{tt}\dot{t}^2+g_{rr}\dot{r}^2+g_{\varphi\varphi}\dot{\varphi}^2=0,
\eeq
which can be rewritten as 

\beq
\dot{r}^2=E^2-V(r).
\eeq
The effective potential $V$ is given by

\beq
V\equiv \frac{L^2}{g_{rr}g_{\varphi\varphi}} + E^2\left( 1+\frac{1}{g_{rr}g_{tt}}\right).
\eeq
In the coordinate $r$, the metric is given by Eq.(\ref{metric1}), and the effective potential is

\beq
V(r) = \frac{L^2r^2}{r^4 + r_{\rm th}^4}
\eeq
The maximum of $V(r)$ is at $V_M = L^2/2r_{\rm th}$.
The following plot gives the effective potential as a function of the $r$ coordinate
for different values of the angular momentum, and a fixed $r_{\rm th}$.

\begin{figure}[h] 
\centerline{\psfig{file=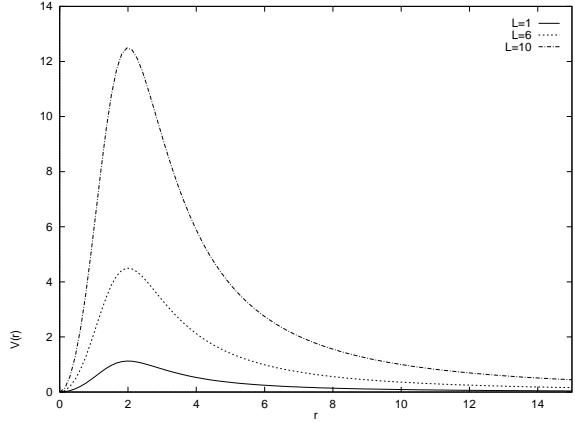,width=8cm,angle=-90}}
\caption{Plot of the effective potential $V$ versus the $r$ coordinate, for $r_{\rm th} =2$.} 
\label{vrho}
\end{figure}
From this plot we see that photons coming from $r=\infty$  with 
energy smaller than $V_M$ are reflected by the potential barrier. Hence they turn back without reaching the throat. Photons with $E>V_M$ will inevitably pass through the throat and will not turn back.

The motion of the photons can also be described from the
plot of the effective potential in the $\rho_\pm$ coordinates. Let us recall first what happens in an ultrastatic spherically symmetric gravitational wormhole with the following metric:

\beq
ds^2 = dt^2 - \frac{d\rho^2}{1-\frac{\rho_{\rm th}^2}{\rho^2}}-\rho^2\;d\Omega^2 .
\label{whmet}
\eeq
In this case the effective potential is given by
\beq
V = \left( 1 - \frac{\rho_{\rm th}^2}{\rho^2}\right) \left(\frac{L^2}{\rho^2} - E^2\right) + E^2 .
\nonumber
\eeq
From this expression, it is easily seen that only particles for which $L^2/E^2<\rho_{\rm th}^2$ will reach the throat. This feature is displayed in the next plot.

\begin{figure}[h] 
\centerline{\psfig{file=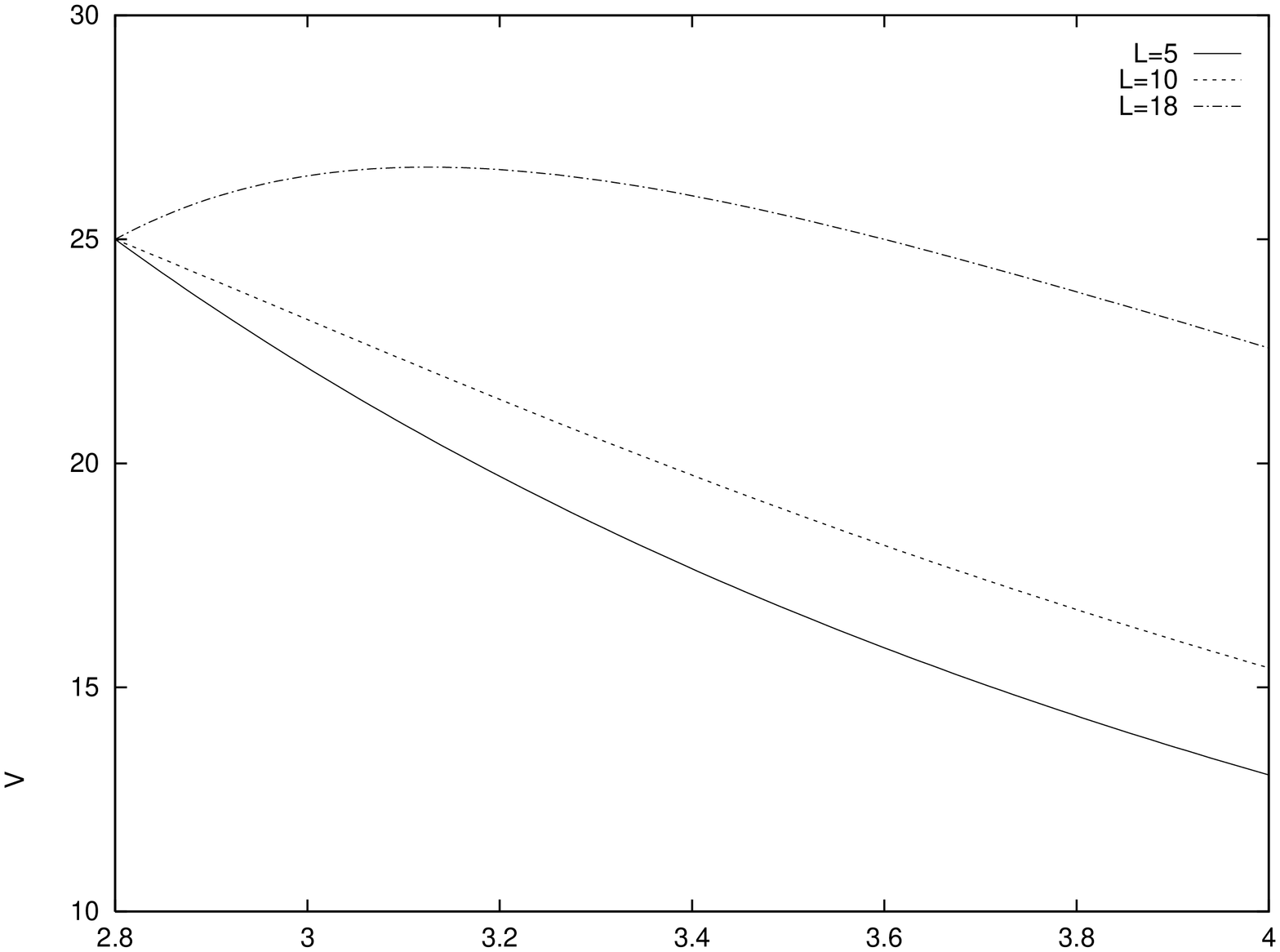,width=8cm,angle=-90}}
\caption{Plot of the effective potential $V$ versus the $\rho$ coordinate, for $\rho_{\rm th} =2.8$, and $E=5$.} 
\label{vrhost}
\end{figure}
Because the metric given in Eq.(\ref{whmet}) describes a symmetric wormhole, after passing through the throat the photon will feel the ``mirror image'' of this effective potential. 

In our case, the expression for $V$ in the $\rho_\pm$ coordinates is
\beq
V(\rho_\pm) = 2\left( \frac{L^2}{\rho_\pm^2} -E^2\right) \frac{\rho_\pm^4 -\rho_{\rm th}^4}{\rho_\pm^2 (\rho_\pm^2 \mp\sqrt{\rho_\pm^4-\rho_{\rm th}^4})} + E^2
\nonumber
\eeq
The following plots give the ``$+$'' and ``$-$'' parts of the effective potential as a function of the $\rho_\pm$ coordinates, for a fixed value of the energy.

\begin{figure}[h] 
\centerline{\psfig{file=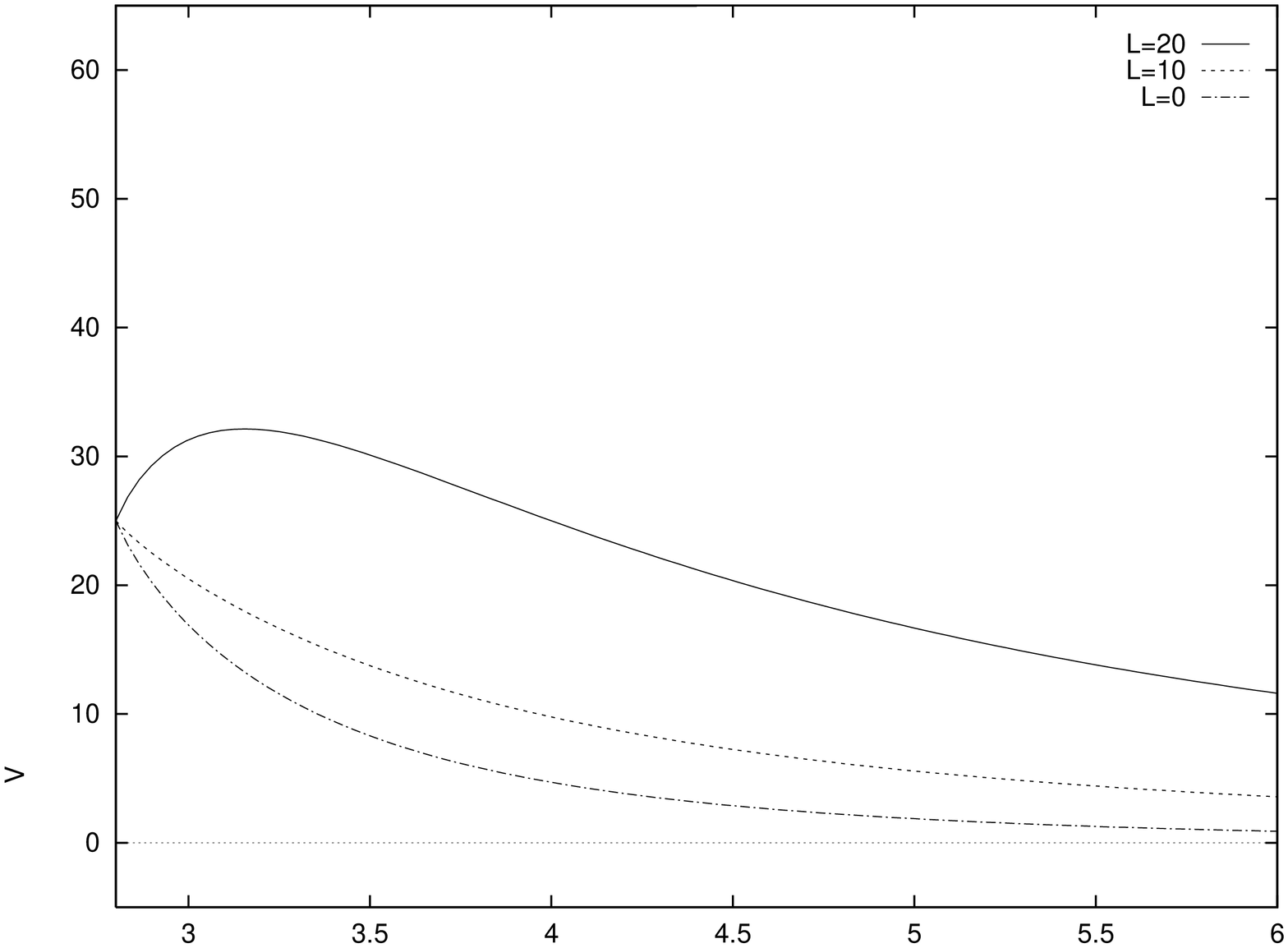,width=8cm,angle=-90}}
\caption{Plot of the effective potential $V$ versus the $\rho_+$ coordinate,
for $E=5$ and $\rho_{\rm th} = 2.8$.} 
\label{vr1}
\end{figure}

\begin{figure}
\centerline{\psfig{file=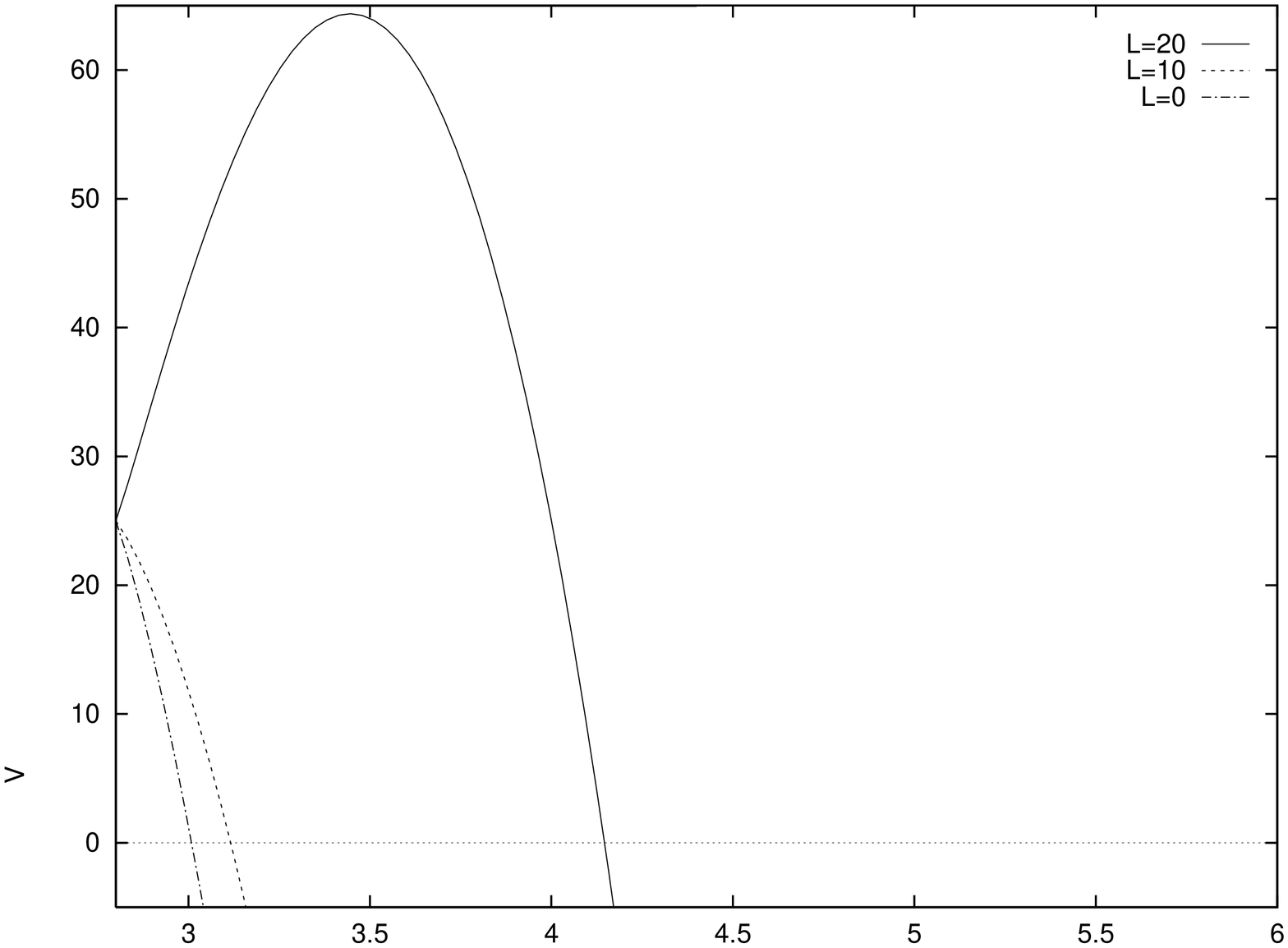,width=8cm,angle=-90}}
\caption{Plot of the effective potential $V$ versus the $\rho_-$ coordinate,
for $E=5$ and $\rho_{\rm th} = 2.8$.} 
\label{vr2}
\end{figure}
Because of the abovementioned asymmetry, here we need both Figs.(\ref{vr1}) and (\ref{vr2}). All the curves in these two figures intersect at $\rho = \rho_{\rm th}$, which corresponds to $V = E^2 = 25$. 
In these two plots (as well in Fig.(\ref{vrhost})) the curves depend on the energy of the photon. This is illustrated in the subsequent plot.

\begin{figure}[h] 
\centerline{\psfig{file=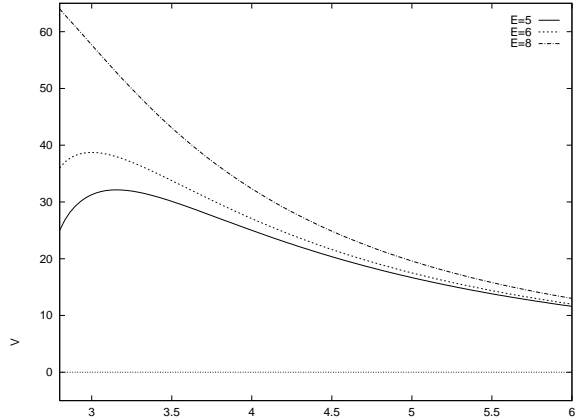,width=8cm,angle=-90}}
\caption{Plot of the effective potential in the $\rho_+$ coordinate for $L=20$ and different values of the energy $E$.}
\label{vr3}
\end{figure}

Comparing Fig.(\ref{vrhost}) with Figs.(\ref{vr1}) and (\ref{vr2}) we see that a photon travelling in the effective metric given by Eq.(\ref{awhmet}) will be under the influence of the same type of effective potential as a photon travelling through a gravitational wormhole. Let us remark that because the path of the photons is invariant under conformal transformations, the wormhole discussed here does not depend on the election of the frame in which the metric is given.

The photons that go from the ``$+$'' patch to the ``$-$'' patch traverse the throat and inevitably escape to $\rho_-\rightarrow\infty$ (see Figs.(\ref{vr1}) and (\ref{vr2})). This motion takes place in the interval $(0,r_{\rm th})$ of the $r$ coordinate (see Fig.(\ref{vrho})). 

\section{Conclusion}

We have shown how a geometrical structure considered previously in a purely gravitational context can also arise in a nonlinear electromagnetic theory in a flat background. We would like to point out once more that this {\em effective wormhole} affects solely the motion of the photons; all other types of matter move according to the flat spacetime dynamical laws. Although we have worked out in detail an example based in Born-Infeld theory, in principle {\em any} nonlinear electromagnetic theory could generate analogues of gravitational structures in a given background. This statement also applies to nonabelian gauge theories with nonlinear dynamics.

Let us remark that the gravitational wormhole requires matter that violates the null energy condition as a source of Einstein's equations. In the case analyzed here, the wormhole is generated through an electromagnetic field that obeys a nonlinear dynamics, and is linked to the metric via Eq.(\ref{geffec}). 

Finally, it should be stressed that the analogy between nonlinear EM and Maxwell's EM in the presence of a nonlinear dielectric medium may open the door to an actual realization of this wormhole (and also of any other of these effective gravitational-like structures) in the laboratory. We will tackle this problem in a forthcoming article. 

\acknowledgements
FB would like to thank the members of the {\em Gravitation and Cosmology} group (LAFEX-CBPF) for their hospitality. SEPB would like to acknowledge  financial support from CONICET (Argentina).  

\section*{appendix}
The expression given in Eq.(\ref{geffec})

\begin{eqnarray}
g^{\mu\nu} = {\cal L}_{F}\,\gamma^{\mu\nu}  - 
4\, {\cal L}_{FF} \,{F^{\mu}}_{\alpha} \,F^{\alpha\nu}
,
\nonumber
\end{eqnarray}
may be viewed as a system of ten equations with the EM field components as unknowns. If we restrict our attention to diagonal $\gamma_{\mu\nu}$ and $g_{\mu\nu}$, the equations

\beq
4\, {\cal L}_{FF} \,{F^{\mu}}_{\alpha} \,F^{\alpha\nu} = 0, ~~~~~~~\mu\neq\nu,
\label{nondiag}
\eeq
impose several constraints on the degrees of freedom of the problem. It turns out that non-trivial solutions are obtained when only two components of $F^{\mu\nu}$ are different form zero: one electric and its dual magnetic component \footnote{This situation is not improved when one considers ${\cal L} = {\cal L}(F,F^*)$,
due to the property ${F^\mu}_\lambda F^{*\lambda\nu}= -\frac{1}{4}F^{\alpha\beta}F^*_{\alpha\beta} \eta^{\mu\nu}$ (see  \cite{nov1}).}. In particular, we can choose a Minkowskian background in spherical coordinates for $\gamma_{\mu\nu}$. Setting
$F^{tr}\equiv A$, $F^{\theta\varphi}\equiv B$, and the rest of the components of $F^{\mu\nu}$ equal to zero  
in order to satisfy Eq.(\ref{nondiag}), we are led to the equations

\begin{eqnarray}
g^{tt}&=&{\cal L}_F-4{\cal L}_{FF} A^2,\nonumber\\
g^{rr}&=&-{\cal L}_F+4{\cal L}_{FF} A^2,\nonumber\\
g^{\theta\theta}&=&-\frac{{\cal L}_F}{r^2}-4{\cal L}_{FF} r^2\sin^2{\theta} \,\,B^2,\nonumber\\
g^{\varphi\varphi}&=&-\frac{{\cal L}_F}{r^2\sin^2{\theta}}-4{\cal L}_{FF} r^2 B^2\nonumber ,
\end{eqnarray}
which is the effective metric given by Eq.(\ref{effmet}).

\end{document}